\documentclass[10pt,journal,cspaper,compsoc]{IEEEtran}
\ifCLASSOPTIONcompsoc
\else
\fi
\ifCLASSINFOpdf
\else
\fi
\usepackage{epsfig}
\usepackage{ifthen}
\usepackage{amsmath}
\usepackage{multicol}
\begin{document}
%
\title{FlexCloud: A Flexible and Extendible Simulator for Performance Evaluation of Virtual Machine Allocation}
%
%
%
\author{ Minxian Xu, Wenhong Tian, Xinyang Wang, Qin Xiong
\thanks{Dr. Tian, Mr. Wang and Miss Xiong are in the school
of Computer Science, University of Electronic Science and Technology of China (UESTC), 610054; Mr. Xu is in the school of software engineering of UESTC.
}}
\maketitle
\begin{abstract}
Cloud Data centers aim to provide reliable, sustainable and scalable services for all kinds of applications. Resource scheduling is one of keys to cloud services. To model and evaluate different scheduling policies and algorithms, we propose FlexCloud, a flexible and scalable simulator that enables users to simulate the process of initializing cloud data centers, allocating virtual machine requests and providing performance evaluation for various scheduling algorithms. FlexCloud can be run on a single computer with JVM to simulate large scale cloud environments with focus on infrastructure as a service; adopts agile design patterns to assure the flexibility and extensibility; models virtual machine migrations which is lack in the existing tools; provides user-friendly interfaces for customized configurations and replaying. Comparing to existing simulators, FlexCloud has combining features for supporting public cloud providers, load-balance and energy-efficiency scheduling. FlexCloud has advantage in computing time and memory consumption to support large-scale simulations. The detailed design of FlexCloud is introduced and performance evaluation is provided.
\end{abstract}
\IEEEpeerreviewmaketitle
\section{Introduction}
With various recent advancements in virtualization, like Grid computing, Web computing, utility computing and related technologies, Cloud computing obtains great development. Cloud computing aims to provide both infrastructure and services on demand through the Internet or intranet \cite{IEEEhowto:Armbrust}, and its benefits can be concluded as hiding and abstraction of complexity, virtualized resources and efficient use of distributed resources. Cloud computing allows the sharing, allocation and aggregation of software, computational and storage network resources on demand. Currently, quite a few IT enterprises products, like Amazon EC2 \cite{IEEEhowto:Amazon}, Google App Engine \cite{IEEEhowto:Google}, IBM blue Cloud \cite{IEEEhowto:IBM} and Microsoft Azure \cite{IEEEhowto:Microsoft} have shown their practice of emerging Cloud computing platforms. Whereas there are many challenging issues to be resolved \cite{IEEEhowto:Armbrust} \cite{IEEEhowto:Beloglazov} \cite{IEEEhowto:Buyya3}, Cloud computing is still considered in its infancy. Youseff et al. \cite{IEEEhowto:Youseff} introduce a detailed ontology of dissecting Cloud into five main layers from top to down: Cloud application (SaaS), Cloud software environment (PaaS), Cloud software infrastructure (IaaS), software kernel and hardware (HaaS), and illustrates their interrelations as well as their inter-dependency on preceding technologies. From structure perspective, Cloud data center can be regarded as a distributed network, containing many computing nodes, storage nodes, or network devices. Each node is composite of a series of resources such as CPU, memory, network bandwidth and so on. In this paper, we focus on Infrastructure as a service (IaaS) in Cloud data centers, and proposing general and flexible definition as well as model that could be used by various cloud providers.

An essential technology in Cloud datacenter is resource scheduling. One challenge problem related to scheduling in Cloud data center is to consider allocation and migration of reconfigurable virtual machines and integrated features of hosting physical machines. Different from existing load-balancing scheduling algorithms that consider only physical servers with one factor such as CPU, the new algorithms treat CPU, memory and network bandwidth integrated for both physical machines (PMs) and virtual machines (VMs). Besides that, real-time virtual machine allocation for multiple parallel jobs and physical machines is taken into consideration. With the development of cloud computing, the size and density of the cloud data center become huge, and problems which need to be solved therewith. For instance, how to manage physical resources and virtual resources intensively and use them dynamically, to improve elasticity and flexibility which can improve service and reduce cost and risk management; and how to help customers build flexible, dynamic, and business growth adapting infrastructure as well as ensure the sustainable development in the future.

Because of the uncertainty of network environments, it is extremely hard to research widely for all these problems in real Internet platform. In addition, the network conditions cannot be predicted or controlled accurately, but affect the validation of strategies. A considerate way in research is developing a simulation system, which supports visualized modeling and simulation in large-scale applications in cloud infrastructure. Data center simulation system can describe the application workload statement, which includes user information, data center position, the amount of users and data centers, and the amount of resources in each data center. Using this information, data center simulation system generates response requests and allocates these requests to virtual machines. By using data center scheduling simulation system, researchers can evaluate suitable strategies such as distributing reasonable data center resources, selecting data center to match special requirements, reducing costs, finding efficient scheduling algorithms and so on.

The major contributions of this paper are as following:
\begin{itemize}
\item
the proposal of a new cloud simulator, FlexCloud, with light weight design to simulate cloud environment;
\item
the design and implementation of a flexible and extendible architecture model that resource, request specification and scheduling algorithms can be easily added;
\item
the validation of the simulator, which has been carried out by comparing realistic data with actual results collected from Lawrence Livermore National Lab \cite{IEEEhowto:ESL} trace;
\item
the performance comparison with CloudSim, which shows FlexCloud has strength in time cost and memory consumption.
\end{itemize}

The remainder parts of this paper are organized as follows: section 2 describes the related work on virtual machine allocation algorithms and compares existing cloud computing simulators from several categories. At the end of this section, the major contributions of this paper are given. Section 3 demonstrates the architectural model of the newly proposed simulator from several aspects, including its layered architecture, scenario, datacenter modeling, VM requests modeling, scheduling algorithms modeling, implemented performance metrics, VM migration modeling, scheduling process modeling and etc. Section 4 presents implementation details and design patterns adopted in FlexCloud. Section 5 and section 6 demonstrate the validation and evaluation of FlexCloud respectively. Finally, this paper ends with the brief conclusions and a discussion on future work.
\section{Related works}
A mount of research has been conducted in resource scheduling algorithms, which are significant for cloud data centers. Mastroaianni \cite{IEEEhowto:Mastroianni} et al. present a self-organizing and adaptive approach for the consolidation of VMs on CPU and RAM resources. Wood et al. \cite{IEEEhowto:Wood} introduce techniques for virtual machine migration and propose some migration algorithms. Zhang et al. \cite{IEEEhowto:Zhang} compare major load balancing scheduling algorithms for traditional web servers. Singh et al. \cite{IEEEhowto:Singh} propose a novel load balancing algorithm called VectorDot for handling the hierarchical and multi-dimensional resource constraints by considering both servers and storage in Cloud computing. Doyle et al. \cite{IEEEhowto:Doyle} propose a system named Stratus to determine the routine decisions for data center requests.

Buyya et al. introduce GridSim \cite{IEEEhowto:Buyya} toolkit for modeling and simulation of distributed resource management for grid computing. Dumitrescu and Foster \cite{IEEEhowto:Dumitrescu} introduce GangSim tool for grid scheduling. Buyya et al. \cite{IEEEhowto:Buyy2} introduce modeling and simulations of Cloud computing environments at application level, a few simple scheduling algorithms such as time-shared and space-shared are discussed and compared. CloudSim \cite{IEEEhowto:Buyy2} is one of Cloud computing simulators, which provides: modeling large-scale cloud computing infrastructure; models for the data center, service agency, scheduling and distributing strategies; virtual engines, which is helpful to create and manage several independent and collaborative virtual services in a data center node; switching flexibly between processing cores with space-sharing and time-sharing. CloudAnalyst \cite{IEEEhowto:Wickremasinghe} aims to achieve the optimal scheduling among user groups and data centers based on the current configuration. Both CloudSim and CloudAnalyst are based on SimJava \cite{Howell} and GridSim \cite{IEEEhowto:Buyya}. Also CloudSim and CloudAnalyst treat a Cloud data center as a large resource pool and consider only application-level workloads, may not suitable for Infrastructure as a service (IaaS) simulation where each virtual machine as resource is considered to be requested and allocated. A CloudSim-based simulation tool considering DVFS energy model is proposed in \cite{IEEEhowto:Guerout}. Kliazovich et al. propose an energy-aware simulation environment named GreenCloud for Cloud datacenters \cite{IEEEhowto:Kliazovich}. Nunez et al. \cite{IEEEhowto:Alberto} introduce a new simulator of cloud infrastructure named iCanCloud using C++ and compare the performance with CloudSim.
\begin{table*}
\scriptsize
\caption{Summary of Cloud Simulators}
\begin{center}
\begin{tabular}{l|l|l|l|l|l}
\hline Items & CloudSim & MDCSim& GreenCloud & iCanCloud & FlexCloud
\\\hline
\hline Platform & any & CSIM & NS2 & OMNET, MPI & any \\
\hline Programming Language & Java & C++/Java &C++/OTcl & C++ & Java \\
\hline Availability  &Open Source & Commercial&Open Source&Open Source & Open Source \\
\hline Graphical Support & Limited (Via CloudAnalyst) &None & Limited (Via Nam)&Full&Full\\
\hline Physical Models& None &None &Limited (Via Plug-in)&Full&Full\\
\hline Models for public cloud & None &None&None&Amazon&Amazon\\
\hline Support for Parallel experiments & No &No &No&Yes&No \\
\hline Support for Energy Consumption Model&Yes&Yes&Yes&No&Yes\\
\hline Support for Migration algorithms &Yes&No&No&No&Yes\\
\hline
\end{tabular} \\
\end{center}
\end{table*}

 Table 1 shows the comparison of some state-of-art cloud simulators as well as FlexCloud proposed in this paper. We compare these cloud simulators from several categories.

Platform: CloudSim and FlexCloud are both implemented with Java, so they can be executed on any machine installed JVM. Built in GridSim and SimJava, CloudSim is heavy to execute. MDCSim is written in CSIM, as for GreenCloud and iCanCloud, they are based on NS2 and OMNET respectively.

Language: The languages implemented with the simulators are related to the platforms. CloudSim and FlexCloud are implemented with Java, MDCSim can be implemented with C++ and Java, and GreenCloud needs combining C++ and OTcl, which is difficult for developers.

Availability: Only MDCSim is commercial, and other four simulators are free or open-source. FlexCloud can be fetched from \cite{FlexCloud}.

Graphical support: MDCSim doesn¡¯t support interface operations. The original CloudSim support no graphical interface, but with CloudAnalyst, the graphical interface are supported. However, full support is not provided in CloudAnalyst, in a whole scheduling process, only the configurations and results can be presented. So we label it ¡°limited¡±, the same reason for GreenCloud. FlexCloud and iCanCloud support whole scheduling process to be showed on the interfaces.

Physical models: iCanCloud and FlexCloud provide detailed simulation for physical analogs for the scheduling. GreenCloud needs to use a plug-in to simulate that.

Models for public cloud providers: Both iCanCloud and FlexCloud use the model suggested by Amazon, in which physical machine and virtual machine specifications are pre-defined.

Parallel experiments: Supporting for multiple machines running the experiments together is a main feature of iCanCloud and that feature is under development. As for FlexCloud, we are working to implement that function as well.

Power consumption model: Except for iCanCloud, other four simulators can support power consumption modeling.

Migration algorithm: CloudSim and FlexCloud support migration algorithms, while other 3 simulators haven't supported that.

In our teaching practice in our university, we have adopted CloudSim, a mature simulator, as a teaching tool assisted, but according to the students¡¯ feedback, CloudSim is a bit complex to use and heavy to execute. That complexity is also a feature of iCanCloud. As for MDCSim, a commercial tool, is not appropriate for researching. Apart from that, it¡¯s not easy to use several languages together in GreenCloud, since it is implemented with C++ and OTcl.

The main contribution of FlexCloud lies in that it is implemented with light weight design, flexible to extend as well as easy to start. Besides the benefits for teaching, we also cooperate with a company researching in resource scheduling to boost the functions of FlexCloud under multi-datacenter environment. They would use FlexCloud to explore suitable algorithms for their company applications.

\section{The Architectural Model of FlexCloud}
Fig.1 shows the overview architecture of FlexCloud with layered components. The top layer is Client Layer that provides the interface for user to configure requests properties and have results feedbacks from lower layers. At this layer, a GUI implemented with Java Swing supports user to configure algorithm types, set PM and VM specifications and select scheduling algorithms. After all settings are completed, the defined configurations would be submitted to lower layer and a sequence of scheduling steps would be processed. Comparison diagrams as well as result outputs would be sent as feedback to Client Layer.
\begin{figure*}[ht]
\begin{center}
{\includegraphics [width=0.85\textwidth, angle=-0] {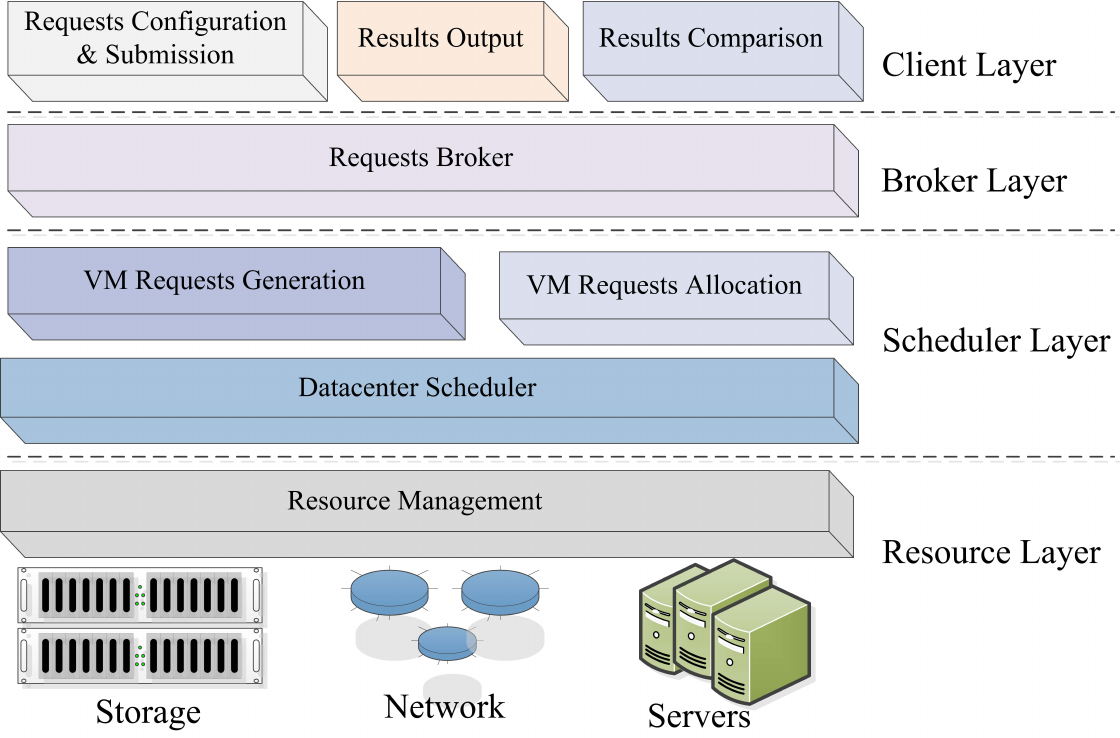}}
\caption{Layered FlexCloud architecture}
\end{center}
\end{figure*}
At lower layer, a Requests Broker is implemented at Broker Layer acting as a mediator between Client Layer and Scheduler Layer. This Layer is responsible for verifying the inputs from Client Layer and transforming the settings into recognized commands at Scheduler Layer. For instance, the number of VM requests submitted from Client Layer would be written into a configuration file, which could be read in the process of scheduling at Scheduler Layer.
Scheduler Layer implements the core functions for FlexCloud system. At this layer, the scheduling process is defined: VM Requests Generation component generates the VM requests with configured properties on user interface; Datacenter Scheduler component schedules the particular algorithms to allocate VMs to corresponding PM according to algorithms; VM Requests Allocation component manages the allocated VMs, including checking the allocation conditions and removing VMs at the end of their lifecycles.
At bottom layer, Resource Layer contains a Resource Management component providing resource that VM requests require and supporting services for higher levels. Besides the component, the physical resource, such as servers, network and storage are resources of the whole system.
\begin{figure*}[ht]
\begin{center}
{\includegraphics [width=0.85\textwidth, angle=-0] {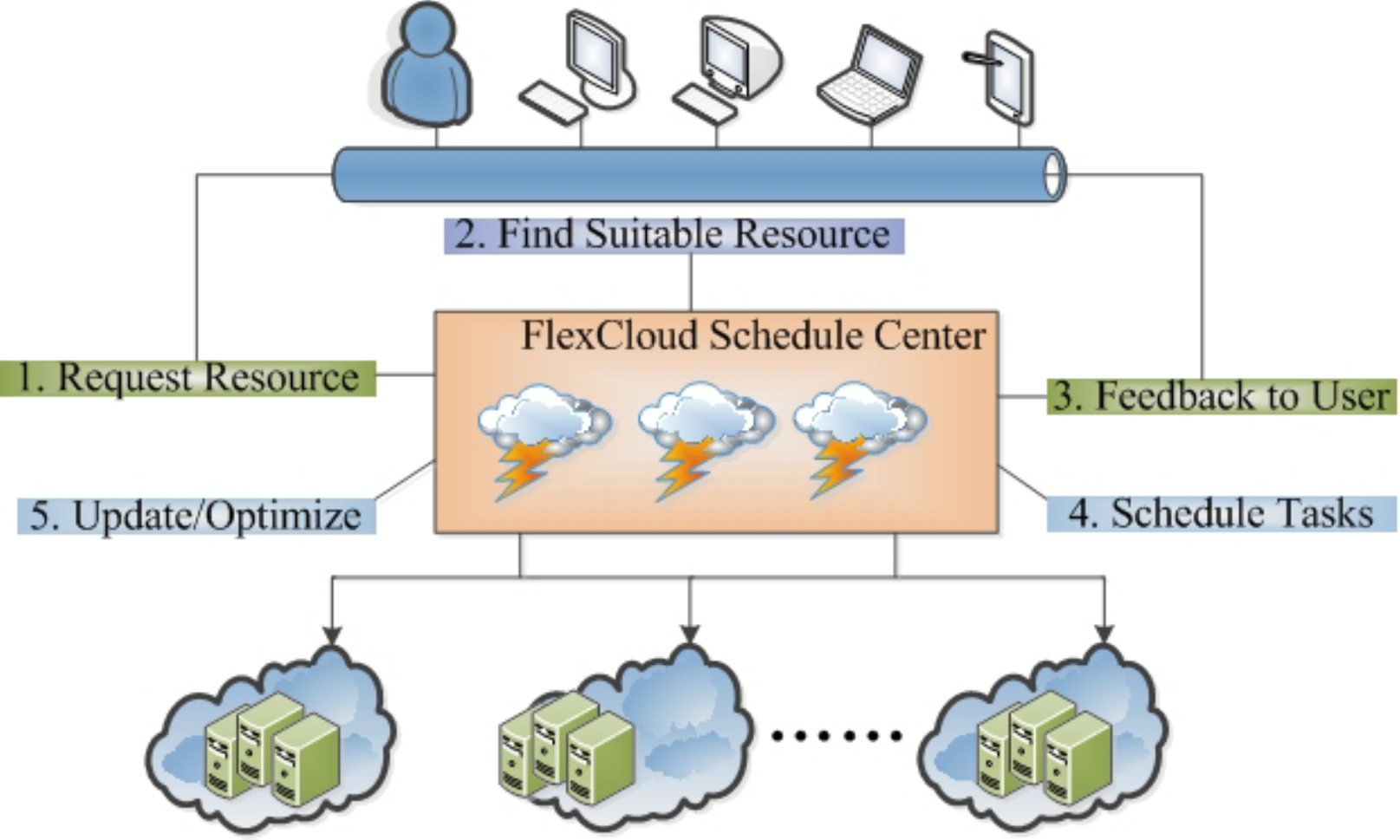}}
\caption{A scenario architecture with FlexCloud}
\end{center}
\end{figure*}

Fig.2 shows an application scenario with FlexCloud. This figure shows the three main components: user, FlexCloud scheduler center and other computing centers. The FlexCloud scheduler center is responsible for the following main tasks: (1) accepting the VM requests sent by users; (2) managing computing centers that in service; (3) finding available computing unit to allocate requests; (4) sending feedback information to users. Computing centers represent a pool of Physical Machines (PMs) or Virtual Machines (VMs), each one configured with a pre-defined specification such as CPU, memory and storage. Users are represented as component that submits a set of jobs to be allocated to specific PM in computing center. These submissions submitted directly to the FlexCloud scheduler center. Then, the requests are managed by this module to be allocated to specific PM in the corresponding data center. After all requests have been processed, a feedback report would be sent back to the user.
\subsection{Modeling the datacenter in FlexCloud}
From computing resource point of view, a data center consists of a number of physical servers (PMs), network devices, storages and other related equipment. A PM contains several kinds of resource, like CPU, memory, storage and bandwidth, etc.
Before VM requests are coming, the PMs are at the state of turned-on, which means the class of Physical Machine is instantiated in FlexCloud. The number of instances depends on the number of PMs would provide services.
\begin{table}
\caption{3 types of physical machines (PMs) suggested
}
\begin{center}
\begin{tabular}{|l|l|l||l|}
\hline PM Pool Type& Compute Units & Memory& Storage
\\\hline
\hline Type 1 & 16 units & 30GB & 3380GB \\
\hline Type 2 & 52 units & 136GB & 3380GB \\
\hline Type 3 & 40 units & 14GB & 3380GB\\
\hline
\end{tabular} \\
\end{center}
\end{table}

In TABLE 2, the 3 suggested types of heterogeneous PMs in FlexCloud are listed, and the configuration can be dynamically set. The type and property values, like CPU, memory, storage and power, are recorded in a configuration XML file (in Fig. 3), which would be loaded into system. Because these property values are in XML file, modification can be easily done either for exactly value or new added property elements. For instance, if more types of PMs are needed, the pair $<pminfo>type-id<pminfo>$ could be created and other values of this type PM could so also be added. Besides the load balance algorithms, we also implement energy-saving algorithms that contain a new property named power consumption. This property is added in the configuration XML file and corresponding methods are added in class PhysicalMachine. The corresponding methods in class PhysicalMachine are responsible for accessing these property values.
\begin{figure}[ht]
\begin{center}
{\includegraphics [width=0.45\textwidth, angle=-0] {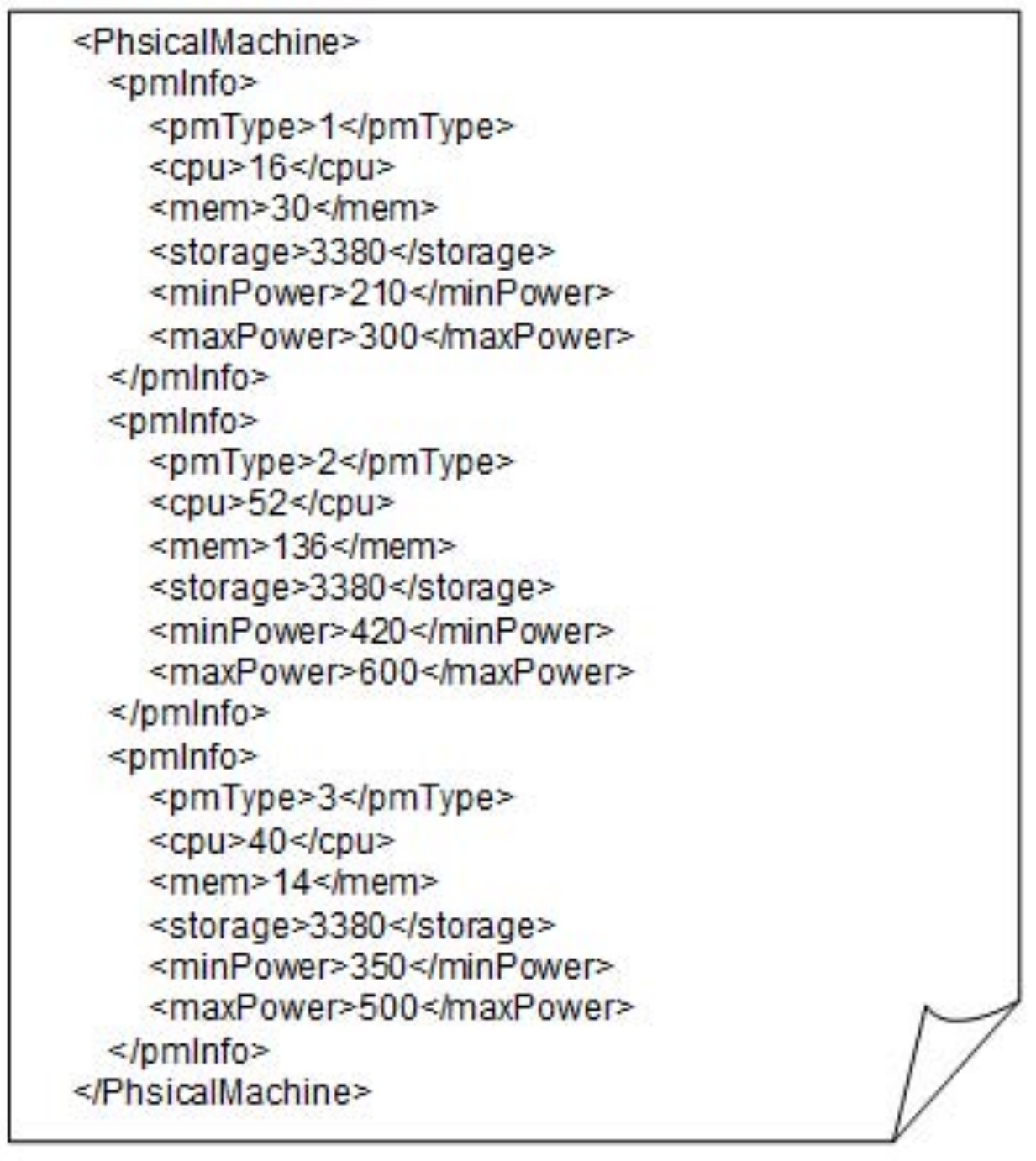}}
\caption{ PM specification in XML file}
\end{center}
\end{figure}

More detailed information related to comparison indices can be found in Section 3.3.
In datacenter model, the left resource capacity decides whether a VM request can be allocated to that PM. At initialization stage, PM has a full capacity resource to offer services. Either the allocation or remove operation would update the available capacity value and influence the later requests allocation.

\subsection{Modeling VM requests in FlexCloud}
We use a simple example to show how VM requests are modeled in FlexCloud in Fig. 4. Slots $\#1, \#2, \ldots, \#6$ represent the time slots in discrete time, which can be treated as a second or a minute that a request is in. For instance, VM2 occupies time slot 3 to 5, so lifecycle of VM2 is 3 slots. The value 0.0625 is proportion of resource occupation, meaning that VM2 would occupy 6.25\% resource of the PM it is allocated to, during time slot 3 to 5. In our model, several VM requests can share the capacity of the same PM at the same time slot only if the capacity is enough.

\begin{figure}[ht]
\begin{center}
{\includegraphics [width=0.45\textwidth, angle=-0] {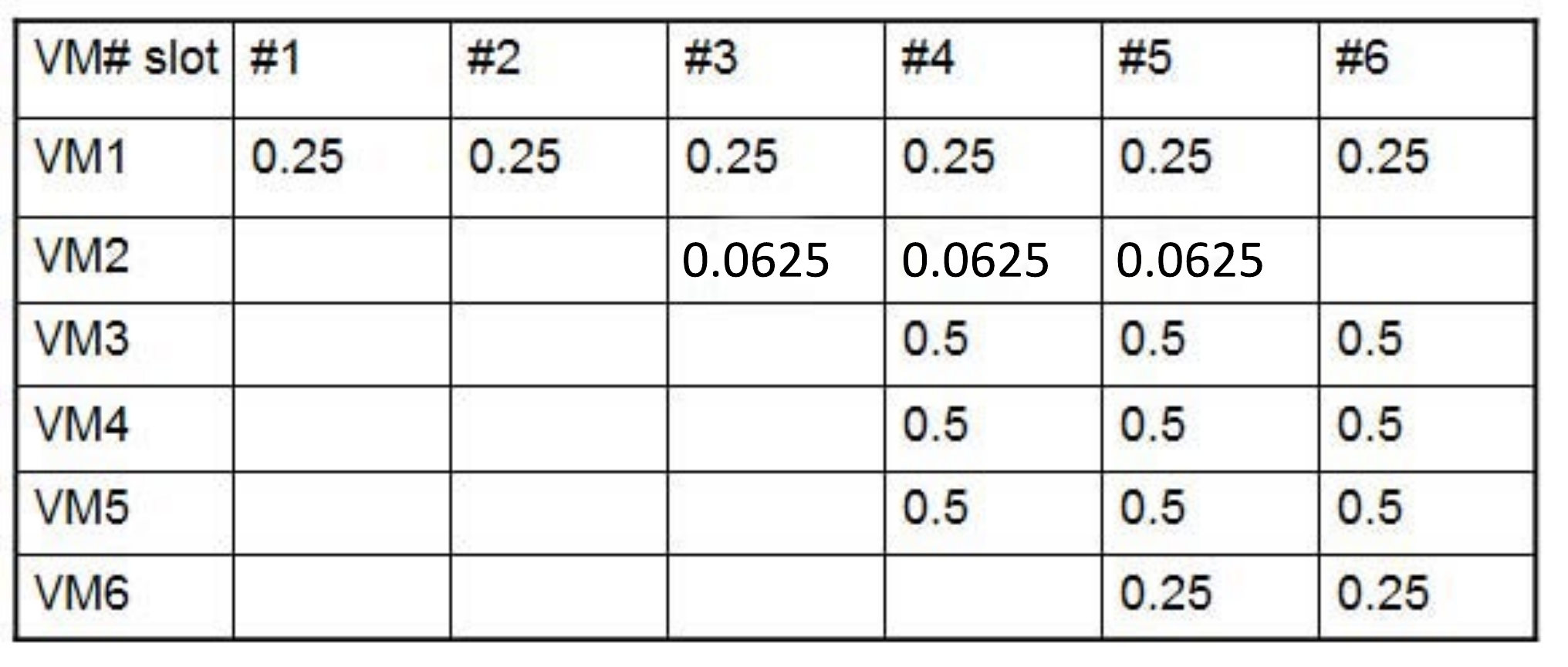}}
\caption{an example of VM requests}
\end{center}
\end{figure}
TABLE 3 shows the corresponding CPU, memory, storage values for different VMs.
Also for extensible reason, these property values are also recorded into a configuration XML file. Once a VM request is allocated to a PM, the left resource capacity would be decreased by the value of that request, and the capacity is increased back when request is released.

In FlexCloud, several VM requests generation approaches have been implemented, in which requests can be generated in Poisson, Normal and Random distributions. When the specific distribution is selected, the start time or duration of the generated requests would follow the distribution. In section 5, we would show the data collected from different distributions. Moreover, it¡¯s available for FlexCloud to import requests data from file in the \emph{Generate VMs} step (see section 3.6), which means it can be tested under realistic data.
\begin{table}
\caption{8 types of virtual machines (VMs) in Amazon EC2}
\begin{center}
\begin{tabular}{|l|l|l||l|}
\hline Compute Units& Memory & Storage& VM Type
\\\hline
\hline 1 units & 1.7GB & 160GB & 1-1(1) \\
\hline 4 units & 7.5GB & 850GB & 1-2(2) \\
\hline 8 units & 15GB & 1690GB&1-3(3) \\
\hline 6.5 units& 17.1GB & 420GB &2-1(4)\\
\hline 13 units & 34.2GB & 850GB &2-2(5)\\
\hline 26 units & 68.4GB & 1690GB &2-3(6)\\
\hline 5 units & 1.7GB & 350GB &3-1(7) \\
\hline 20 units & 7GB & 1690GB &3-2(8)\\
\hline
\end{tabular} \\
\end{center}
\end{table}

\subsection{Modeling Scheduling Algorithms in FlexCloud}
Four kinds of scheduling algorithms are provided in FlexCloud based on scheduling goals and request types. For request types, scheduling algorithms can be divided into online algorithms and offline algorithms, the difference lies in whether the requests information is all known before scheduling. Requests would come and be operated one by one in online algorithm, while requests sequence can be adjusted by processing time or end time because all requests information have been collected before scheduling in offline algorithms. Another division principle is via goal: we consider load balancing and energy saving in FlexCloud.

When comparing the effects of different algorithms, the scheduling process would be same except that the scheduling algorithms are different. For online load balancing comparison, Random, Round-Robin (Round), List Scheduling (LS) algorithms have been implemented. Under the layered architectural model and related design pattern (introduced in later section), new created algorithms can be added to scheduling algorithm library, without influencing other existed algorithms.

We take the Random algorithm, one of the simplest algorithms, as an example to show scheduling algorithm could be modeled, mapped and extended in FlexCloud. Fig.5 shows the pseudo-code of Random algorithm. After a data center scheduler has initialized the PMs and VM requests, Random algorithm would randomly generate an index in the range of 0 and $M$-1 (line 3) for PMs. Then VM request will be allocated to the PM with generated index (line 5), if allocation is successful by checking whether the PM has available resource, then allocated PM needs updating its left capacity (line 7), another index would be generated if allocation is failed and the VM request should be allocated again with a new index (line 8-9). As VM requests have lifecycles, the VM requests should be released from hosting PM to prepare for other VM requests' allocation (line 11 to 12) after their end-time expired. This example is based on singe data center, while under multiple data centers, the index generation process would be involved with data center id generation, rack id generation and PM id generation rather than only PM id generation.
\begin{figure}[ht]
\begin{center}
{\includegraphics [width=0.48\textwidth, angle=-0] {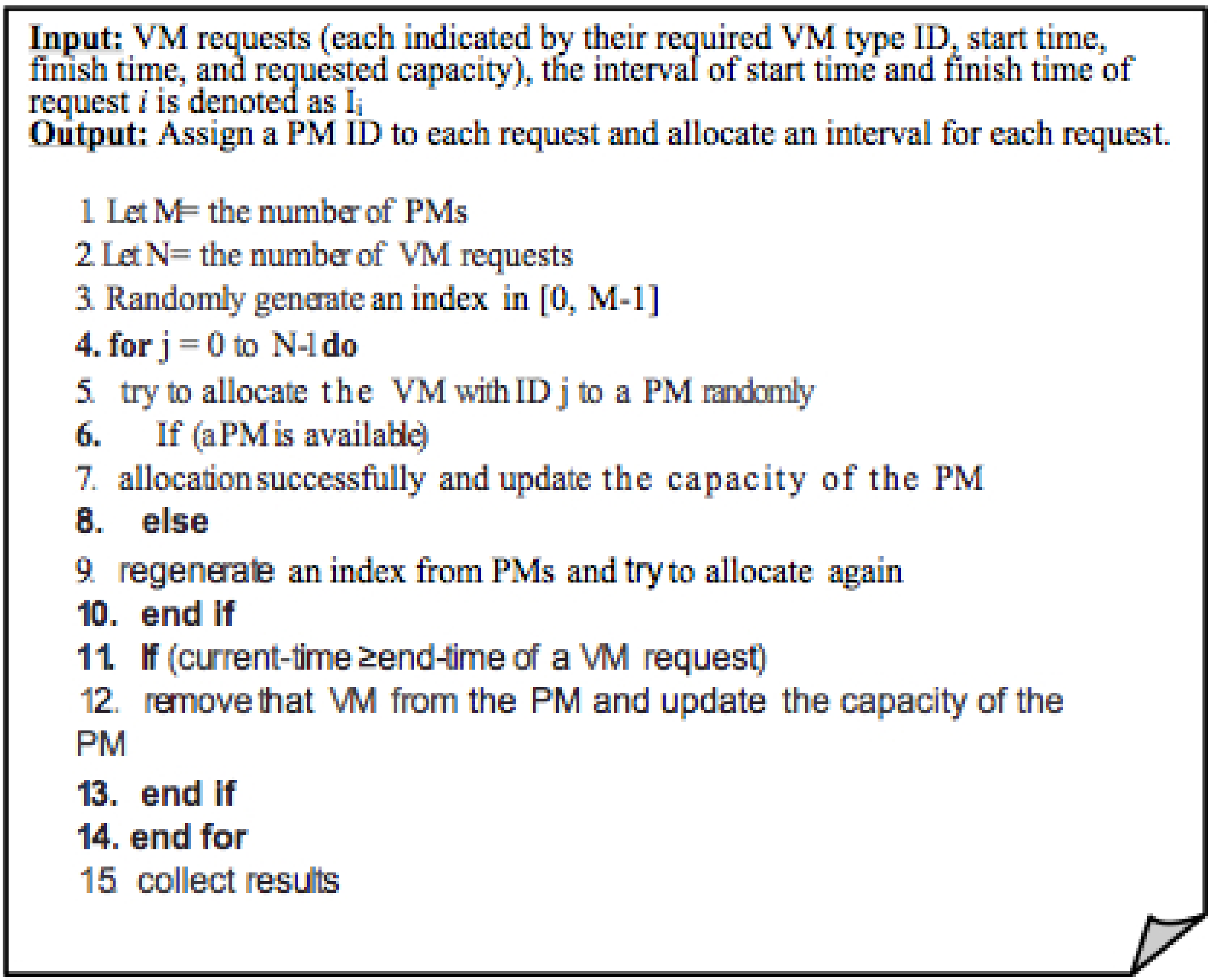}}
\caption{The pseudocode of Random algorithm}
\end{center}
\end{figure}

The algorithm process of R-R and LS is quite similar to Random except the way the index is generated for PMs or VMs. In R-R algorithm, index generation is in a round robin way while the index refers to the PM with the least average utilization in LS algorithm. The more complex algorithms, Post Migration algorithms and Prepartition Algorithm, that considering migrations operations introduced in section 3.5 could also be modeled based on these principles. As for offline load balancing algorithms, a procedure of request processing should be added before line 4. The processing procedure may change the requests order by processing time or end time or other features. Also, the process of online/offline energy-saving algorithms taken is similar to online/offline load balancing algorithms.
\subsection{Performance Metrics in FlexCloud}
In this section, we introduce the major performance metrics we used in FlexCloud:

For load balancing algorithms:\\
Average utilization: Each PM would have the utilization value in scheduling process, and average utilization is the arithmetic average value of all PMs in the data center;\\
PM resource: $PM_i(i,PCPU_i,PMem_i,PStorage_i)$, $i$ is the index number of PM, $PCPU_i,PMem_i,PStorage_i$ are the CPU, memory, storage capacity of that a PM can provide.\\
VM resource: \\$VM_j(j,VCPU_j,VMem_j,VStorage_j,T_j^{start},T_j^{end})$, $j$ is the VM type ID, $VCPU_j,VMem_j,VStorage_j$ are the CPU, memory, storage requirements of $VM_j$, $T_j^{start},T_j^{end}$ are the start time and end time, which are used to represent the life cycle of a VM.\\
Time slot: we consider a time span from 0 to T be divided into parts with same length. Then $n$ parts can be defined as $[(t_1-t_0),(t_2-t_1),\ldots,(t_n-t_{n-1})]$, each time slot $T_k$ means the time span $(t_k-t_{k-1})$.\\
Average CPU utilization of $PM_i$ during slot 0 and $T_n$:
\begin{equation}
PCPU_i^U=\frac{\sum_{k=0}^{n} (PCPU_i^{T_k}\times T_k)}{\sum_{k=0}^{n} T_k}
\end{equation}
And memory $PMem_i^U$ and storage $PStorage_i^U$ utilization of both PMs and VMs can be computed in the same way. Similarly, average CPU utilization of a VM can be computed.\\
Integrated load imbalance value $ILB_i$ of $PM_i$. The variance is widely used as a measure of how far a set of values is spread out from each other in statistics. Using variance, an integrated load imbalancing value $ILB_i$ of server $i$ is defined
\begin{align}
ILB_i =& \frac{(Avg_i-CPU_u^A)^2}{3} +\frac{(Avg_i-Mem_u^A)^2}{3} \nonumber \\
&+\frac{(Avg_i-Storage_u^A)^2}{3}
\end{align}
where
\begin{equation}
Avg_i = \frac{PCPU_i^U+PMem_i^U+PStoarge_i^U}{3}
\end{equation}
and $CPU_u^A,Mem_u^A,Storage_u^A$ are respectively the average utilization of CPU, memory and storage in a Cloud data center.\\
$ILB_i$ is applied to indicate load imbalance level comparing utilization of CPU, memory and network bandwidth of a single server itself.\\
Makespan: is as same as traditional definition, and therefore the capacity$\_$makespan of all PMs can be formulated as below:
\begin{equation}
capacity\_makespan = \max_i{(L_i)}
\end{equation}
Load efficiency (skew of makespan): is defined as the (minimal average load divided by maximal average load) on all machines:
\begin{equation}
skew(makespan) = \frac{\min_i(L_i)}{\max_i(L_i)}
\end{equation}
where $L_i$ is the load of PM $i$. Skew shows the load balancing efficiency to some degree.\\
Capacity$\_$makespan: In any allocation of VM requests to PMs, we can let $A(i)$ denote the set of VM requests allocated to machine $PM_i$, under this allocation, machine $PM_i$ will have total loads,
\begin{equation}
L_i=\sum_{j\in A(i)}c_jt_j
\end{equation}
Based on the above definitions and equations, we have developed another metric, capacity$\_$skew on load balancing algorithm for the new situation as follow:\\
Skew of capacity$\_$makespan is defined as the minimal capacity$\_$makespan over maximal capacity$\_$makespan on all machines (referring to equation (6)):
\begin{equation}
skew(capacity\_makespan) = \frac{\min{\sum_{j\in A(i)}c_jt_j}}{\max {\sum_{j\in A(i)}c_jt_j}}
\end{equation}
where $c_j$ is the capacity (for example CPU) requests of $VM_j$ and $t_j$ is the span of request $j$ (i.e., the length of processing time of request $j$).

For energy saving algorithms, following indices are provided:\\
1) The total number of PMs turned-on during the scheduling;\\
2) Rejected number of VM requests: VM requests which cannot be served by the data center resources;\\
3) Total energy consumption: the energy consumption of all PMs (including VMs allocated on them); a less total energy consumption value reflects a better energy saving effect for a given set of requests.

In \cite{IEEEhowto:Beloglazov}, authors found that CPU utilization is typically proportional to the overall system load, and proposed a power model defined in equation (8):
\begin{equation}
P(u) = kP_{max} + (1-k)P_{max}u
\end{equation}
where $P_{max}$ is the maximum power consumed when the server is fully utilized; $k$ is the fraction of power consumed by the idle server (studies show that on average it is about 70\%); and $u$ is the CPU utilization. This paper focuses on CPU power consumption, which accounts for main part of energy comparing to the other resources such as memory, disk storage and network devices. In FlexCloud, we use the power model defined in (8). Equation (8) is further reduced to (9):
\begin{equation}
P = P_{min} + (P_{max} - P_{min})u
\end{equation}
where $P_{min}$ is the power of given PM when its CPU utilization is zero (the PM is idle without any VM running). In real environment, the utilization of the CPU may change over time due to the workload variability. Thus, the CPU utilization is a function of time and is represented as $u(t)$. Therefore, the total energy consumption by a PM ($E_i$) can be defined as an integral value of the power consumption function over a period of time as in (10):
\begin{equation}
E_{i} = \int_{t0}^{t1} P(u(t))dt
\end{equation}
If $u(t)$ is constant over time, for example average utilization is adopted, $u(t) = u$, then $E_i = P(u)\times(t_1 - t_0)$.

The total energy consumption of a cloud data center is computed as (11):
\begin{equation}
E_{DC} = \sum_{i=1}^{n}E_i
\end{equation}
It is the sum of energy consumed by all PMs. Notes that energy consumption of all VMs on PMs is included.

Also confidence intervals can be calculated for different metrics as follows: Let $x_1, x_2, x_3,..., x_n$ be the calculated metrics (such as $IBL_{tot}$ ~and~ $E_{cdc}$ values etc.) from $n$ times of repeated simulations. Then the mean is
\begin{equation}
x_{mean}=\frac{1}{n}\sum_{i=1}^{n} x_i
\end{equation}
and the standard deviation $s$ is
\begin{equation}
s=\sqrt{\frac{\sum_{i=1}^{n} (x_{mean}-x_i)^2}{n-1}}
\end{equation}
and the confidence interval at 95$\%$ confidence is given by
\begin{equation}
(x_{mean}-1.96\frac{s}{\sqrt{n}}, x_{mean}+1.96\frac{s}{\sqrt{n}})
\end{equation}
\\
Above are basic the metrics that already implemented in FlexCloud. Other metrics could also be included for further research.

\subsection{Modeling Virtual Machine Migrations in FlexCloud}
There is lack of virtual machine migration modeling in existing simulation tools. In \cite{Tian2014}, the detailed algorithms about migration are introduced and compared. In this section, we provide brief introduction to virtual machine migration modeling in FlexCloud. The key difference from allocation is that the migration objectives and the choose of resource and destination PMs. Two typical migration algorithms are introduced in FlexCloud:

Post Migration algorithm: Firstly, it processes the requests in the same way as LPT (Longest Processing Time first) does. Then the average capacity$\_$makespan of all jobs is calculated. The up-threshold and low-threshold of the capacity$\_$makespan for the post migration are calculated through the average capacity$\_$makespan multiplied by a factor (in this paper we set the factor as 0.1, so the up-threshold is average capacity$\_$makespan multiplied by 1.1 and the low-threshold is multiplied by 0.9). Off course the factor can be set dynamically to meet different requirements; however, the larger the factor is, the higher imbalance is. A migration list is formed by collecting the VMs taken from PMs with capacity$\_$makespan higher than the low-threshold. The VMs would be taken from a PM only if the operation would not lead the capacity$\_$makespan of the PM to be less than the low threshold. After that, the VMs in the migration list would be re-allocated to a PM with capacity$\_$makespan less than the up-threshold. The VMs would be allocated to a new PM only if the operation would not lead the capacity$\_$makespan of the PM to be higher than the up-threshold. There may be still some VMs left in the list, finally the algorithm allocates the left VMs to the PMs with the lowest capacity$\_$makespan until the list is empty.

Capacity\_makespan Prepartition Algorithm: novel work proposed by ourselves. For a given set of VM reservations, let us consider there are $m$ PMs in a data center and denote OPT as the optimal solution for a given set of $J$ VM reservations. Firstly define
\begin{equation}
P_0=max\{max_{j=1}^{J} {CM_j},\frac{1}{m}\sum_{j=1}^{J} CM_j\}\leq OPT
\end{equation}
$P_0$ is a lower bound on OPT.
The Capacity\_makespan Prepartition algorithm is introduced in detailed in \cite{Tian2014}. It firstly computes balance value by equation (15), defines partition value ($k$) and finds the length of each partition (i.e. $\lceil P_0/k\rceil$, which is the max time length a VM can continuously run on a PM). For each request, Prepartition equally partitions it into multiple $\lceil P_0/k\rceil$ subintervals if its CM is larger than $\lceil P_0/k\rceil$, and then finds a PM with the lowest average capacity$\_$makespan and available capacity, and updates the load on each PM. After all requests are allocated, the algorithm computes the capacity$\_$makespan of each PM and finds total partition (migration) numbers. For practice, the scheduler has to record all possible subintervals and their hosting PMs of each request so that migrations of VMs can be conducted in advance to reduce overheads.\\
FlexCloud therefore can evaluate the performance of different migration algorithms; the evaluation process is similar to allocation algorithms.

\subsection{The Scheduling Process in FlexCloud}
The major steps of the scheduling process in FlexCloud are as followings:\\
1). Booting PMs: it loads the configuration XML file containing PM specifications set by user from user interface. After needed information is collected, instances of PMs are created to prepare for VM allocation.\\
2). Generating traces (VM requests): it loads the configuration XML file containing VM specifications and VM traces from user interfaces.\\
3). Comparing scheduling algorithms: two or more iterators would collect the compared algorithms and compared indices. All selected algorithms will be compared and corresponding indices are collected.\\
4) Output results: the comparison results are outputed in both text or diagrams format.\\
\begin{figure}[ht]
\begin{center}
{\includegraphics [width=0.45\textwidth, angle=-0] {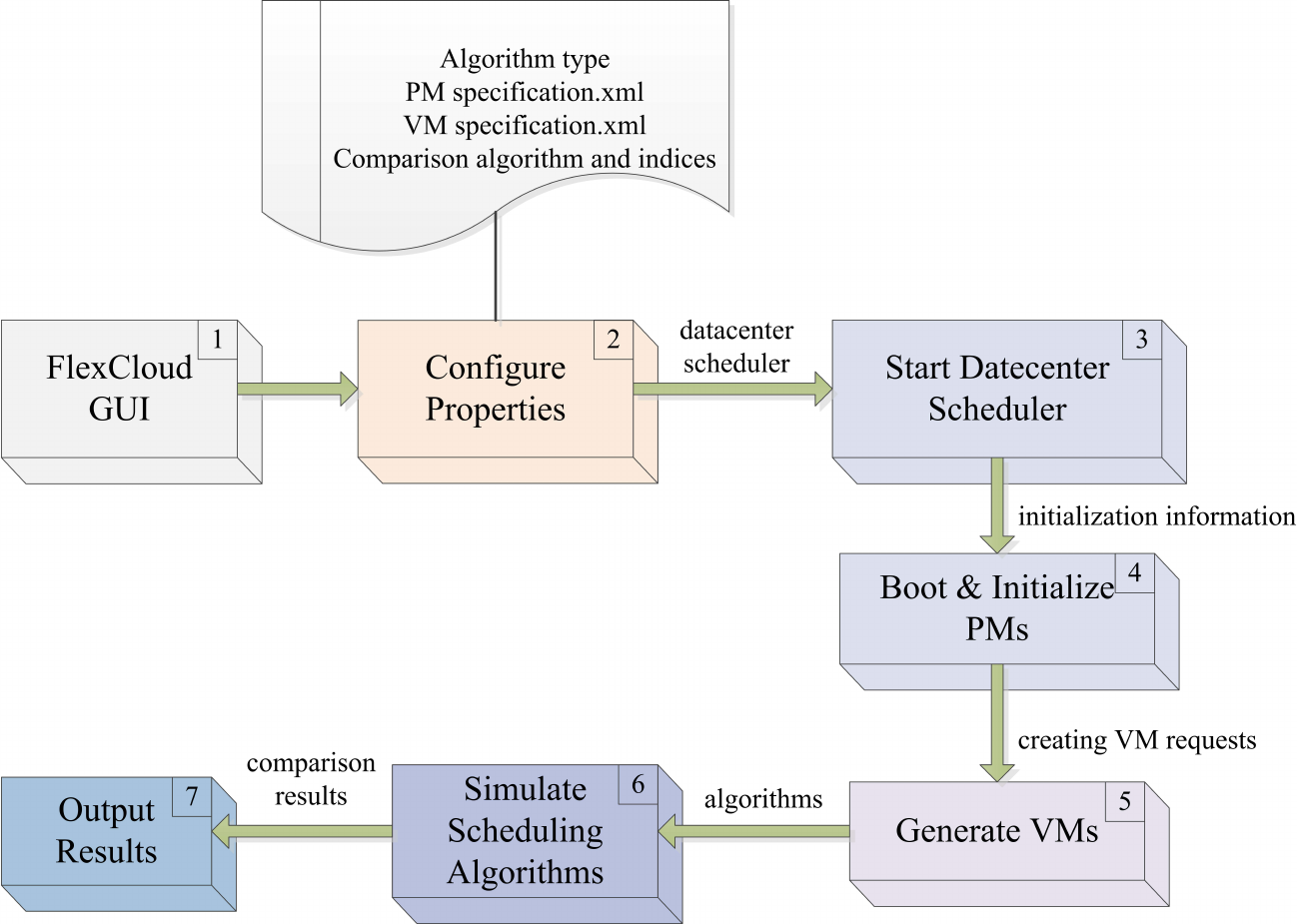}}
\caption{Scheduling process in FlexCloud}
\end{center}
\end{figure}
For building a more flexible system, the scheduling process only defines the basic framework process and customization may be improved based on this process. Before booting PMs, the PM specifications can be modified in configuration file. As for generating VM requests traces, besides the configuration file, more VM requests creation methods can be implemented. Various algorithms can be developed in algorithm scheduling process and other results format may be adopted for better visual effects. Fig. 6 shows a scheduling process combing user interface configurations and basic scheduling process.
\section{ Implementation of FlexCloud}
In this section, we will introduce the detailed implementation of FlexCloud from design patterns' point of view. The design principles are mainly aiming at satisfying agile system goals with flexibility and extendibility.
\subsection{Main Features in FlexCloud}
Considering design principles, FlexCloud mainly has following novel features:\\
(1) FlexCloud is built on Java platform and can be run on a single computer installed JVM to simulate large scale cloud infrastructure as a service (IaaS). A computer with 4 GB memory can simulate larger scale applications. We test the condition when the Java environment would throw an OutOfMemory exception by increasing requests gradually. With a 4 GB memory computer, experiments can simulate scheduling process of more than 100,000 requests. We have extended our tests with computers with 2GB memory, that configuration can simulate requests ranging from 25,000 to 50,000.\\
(2) A user-friendly GUI is provided and lots of customized configurations can be set to satisfy various simulation assumptions. The basic operation includes: select algorithm type, set VM numbers, set average duration, set start time, set total number of PMs, select comparison algorithms and indices. Of course, without GUI, user can also simulate a cloud datacenter and scheduling process in .java class file as well.\\
(3) A scheduling process framework is defined, each step of process can be extended easily in agile style.\\
(4) New scheduling algorithms and performance metrics are flexible and extendable to add in; currently load-balancing and energy-efficiency scheduling algorithms are considered.\\
(5) Virtual machine migration is modeled, this is still lack in current simulation tools.\\
\subsection{Design Patterns in FlexCloud}
\begin{figure}[ht]
\begin{center}
{\includegraphics [width=0.5\textwidth, angle=-0] {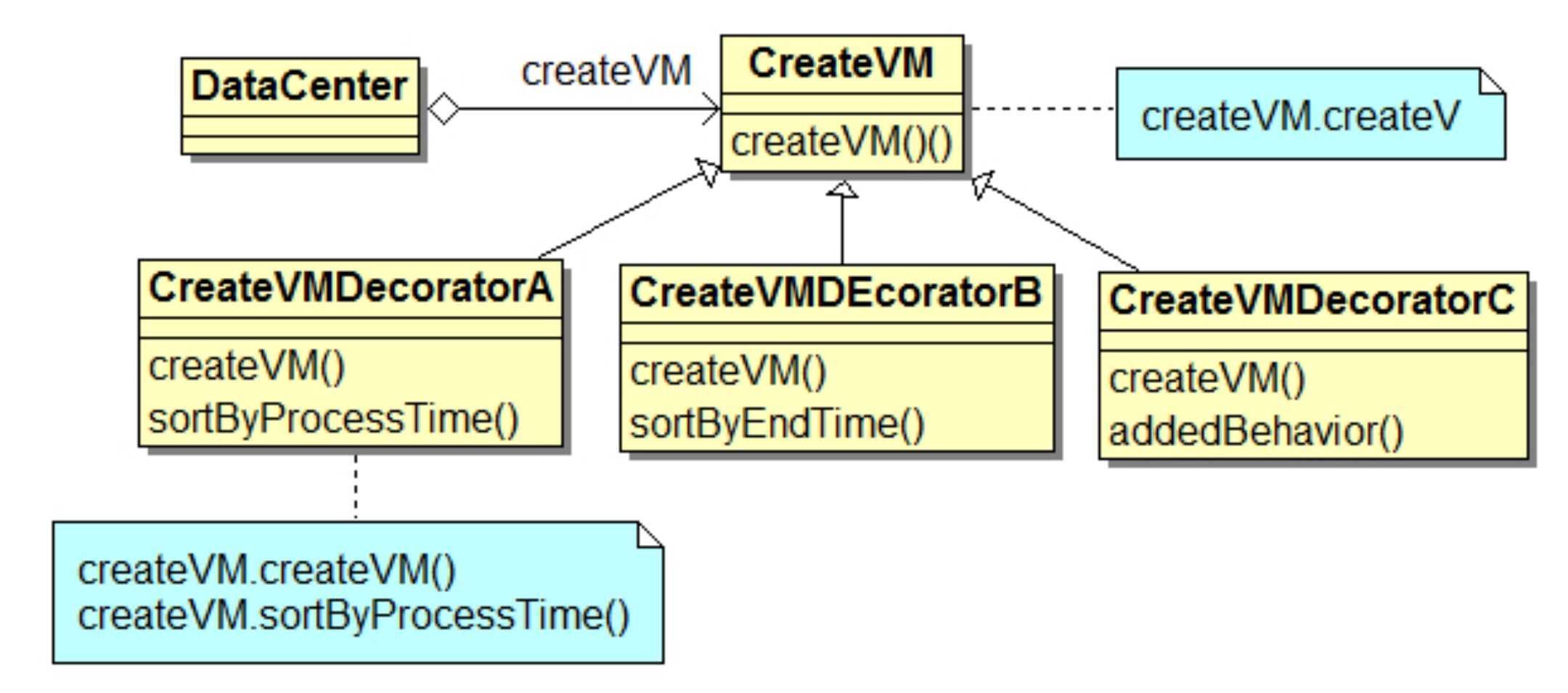}}
\caption{Decorator pattern in FlexCloud}
\end{center}
\end{figure}
FlexCloud has adopted some design patterns to meet the extendible goal.
Fig.7 shows decorator pattern to meet the requirement that requests generation approaches may differ. Class CreateVMDecoratorA and CreateVMDecorateB extend the createVM() method of CreateVM and add a new behavior method. With this pattern, when new requests generation approaches are needed, we can rewrite the method addedBehavior(). Under this method, function of class CreateVM can be dynamically added or deleted. For instance, new resource is needed, resource collection codes can be put in the addedBehavior() method rather than change the existing codes or add new classes.
\begin{figure}[ht]
\begin{center}
{\includegraphics [width=0.5\textwidth, angle=-0] {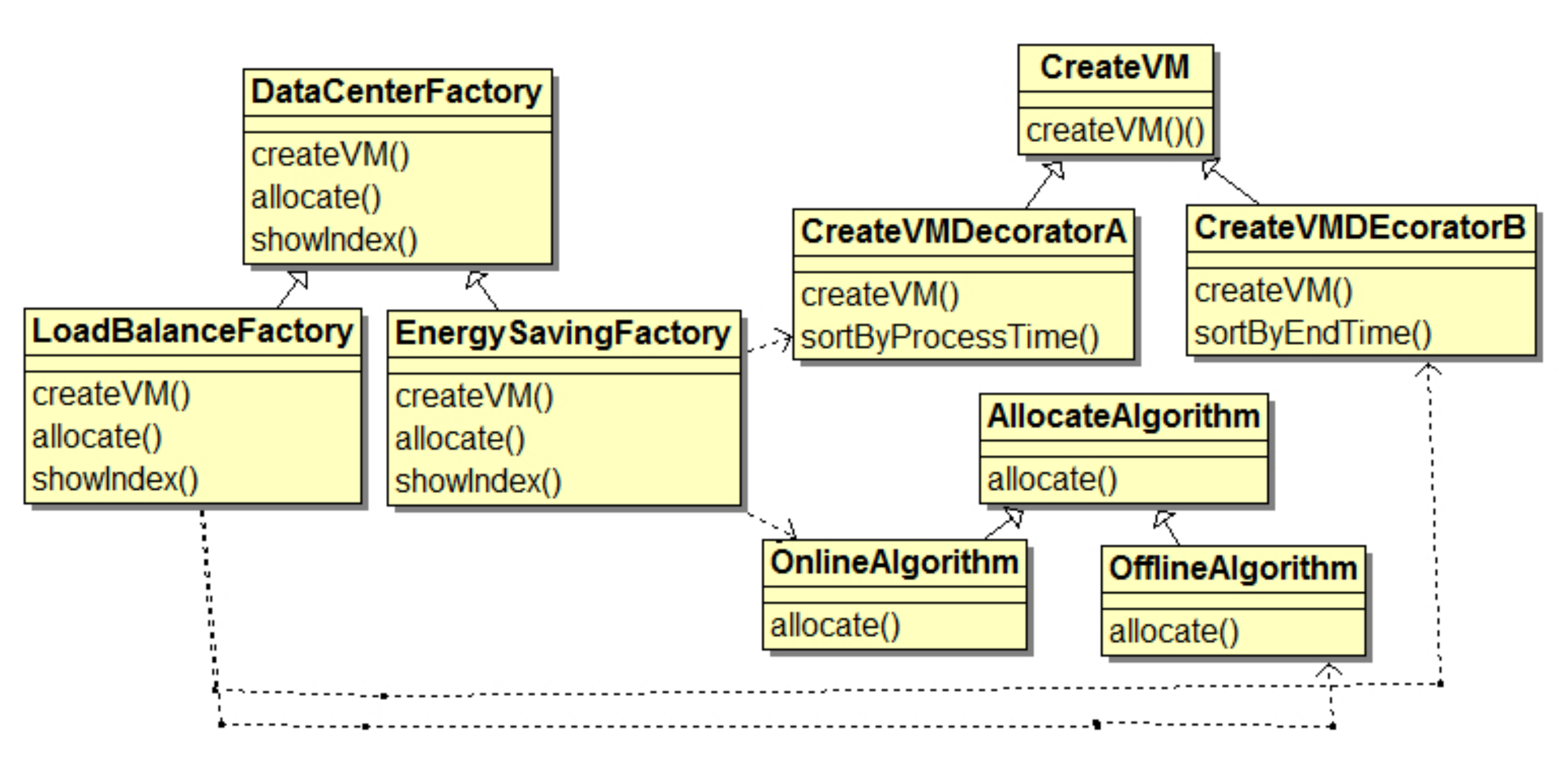}}
\caption{Abstract Factory pattern in FlexCloud}
\end{center}
\end{figure}

Fig.8 shows the Abstract Factory pattern to meet the requirement of different compositions of requests generation approaches and scheduling algorithms. An instance of LoadBalanceFactory would combine a request generation approach from CreateVM and scheduling algorithms from generalization of AllocateAlgorihm. These different combinations can produce diverse scheduling process, like set requests order by processing time and scheduled by a subtype of class OnlineAlgorithm. Adopting this design pattern can avoid fixed composition and gain a better extendable effect.
\begin{figure}[ht]
\begin{center}
{\includegraphics [width=0.5\textwidth, angle=-0] {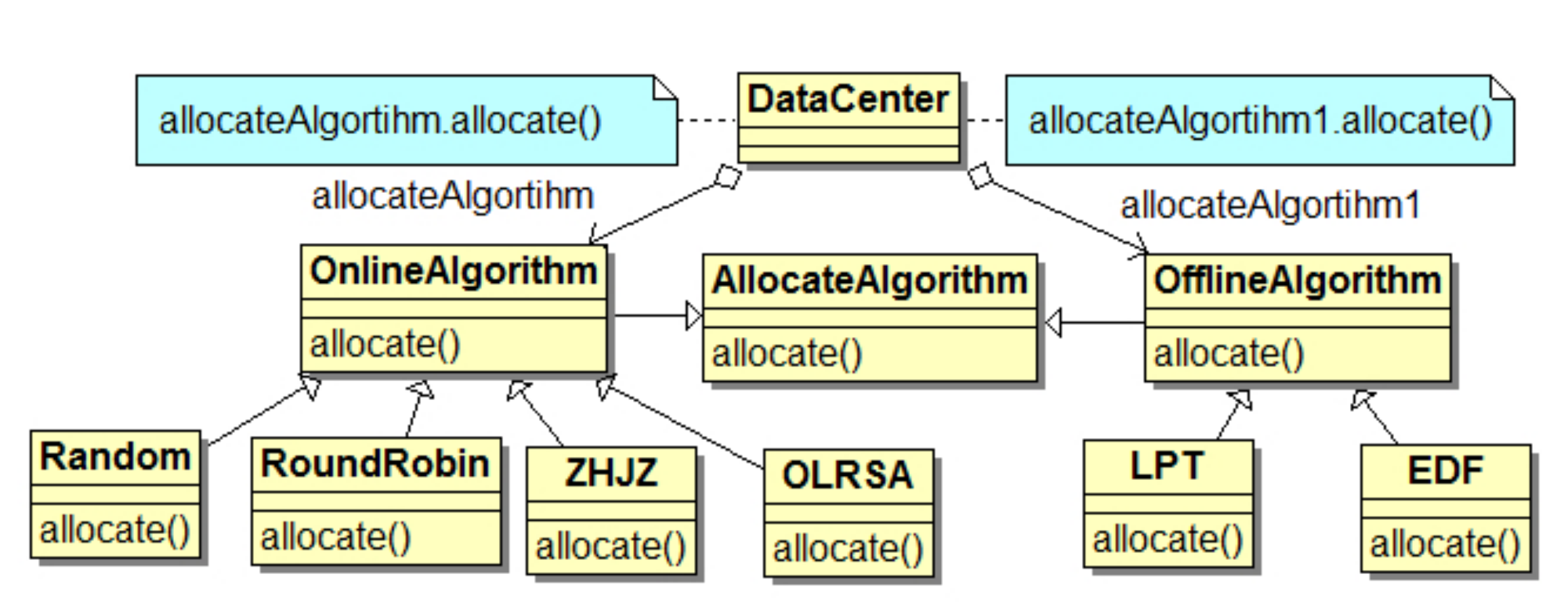}}
\caption{Strategy pattern in FlexCloud}
\end{center}
\end{figure}

Strategy pattern in Fig.9 defines a series encapsulate scheduling algorithm classes: Random, Round-Robin, and OLRSA [18] for OnlineAlgorithm and LPT (Longest Process Time) etc. for OfflineAlgorithm, which can be substituted with each other, enabling scheduling algorithms be independent on the changes from users. With strategy design pattern, when a new algorithm is joined, only allocate() method in new joined algorithm should be implemented. After that, the new joined algorithm can work as same as existing algorithms.
\begin{figure}[ht]
\begin{center}
{\includegraphics [width=0.5\textwidth, angle=-0] {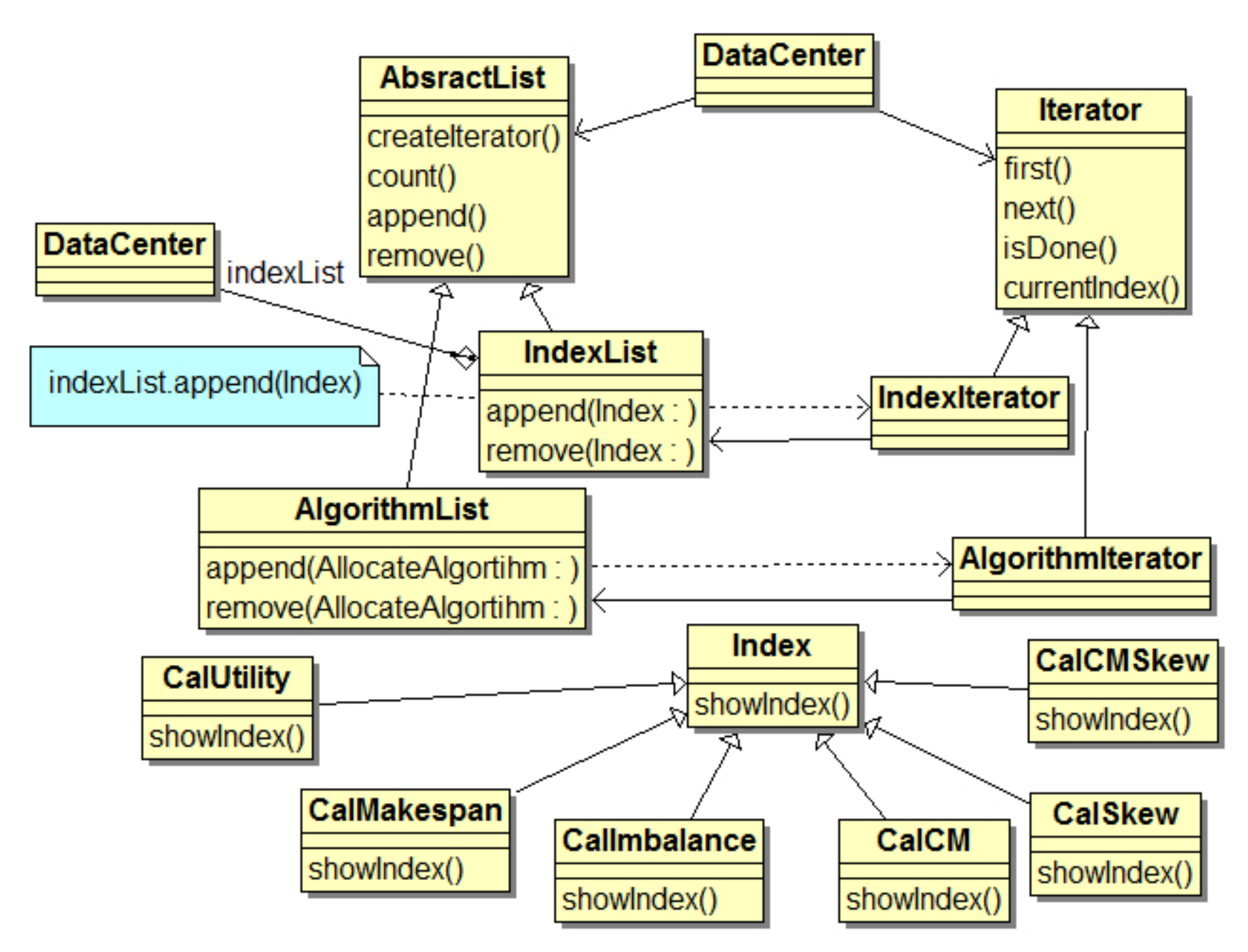}}
\caption{Iterator pattern in FlexCloud}
\end{center}
\end{figure}

Fig.10 shows the Iterator pattern to meet the requirement of algorithm and indices results comparison in data centers. After user have selected the comparison indices and algorithms on user interface, the selected algorithms and indices would be added to separate list, at the same time, Iterator for algorithm and index, Index Iterator and Algorithm Iterator, would be generated. When outputting results, the Iterator would schedule algorithms in Iterator one by one and output indices results with showIndex() method in order. To satisfy the Iterator, both algorithms and indices should extend from their base class. The strength for this design pattern is also a favorable extendibility. It's intelligent when new algorithms and indices are implemented, the Iterator would compare results only if they are appended to it.
\subsection{Communications Between Entities}
\begin{figure}[ht]
\begin{center}
{\includegraphics [width=0.5\textwidth, angle=-0] {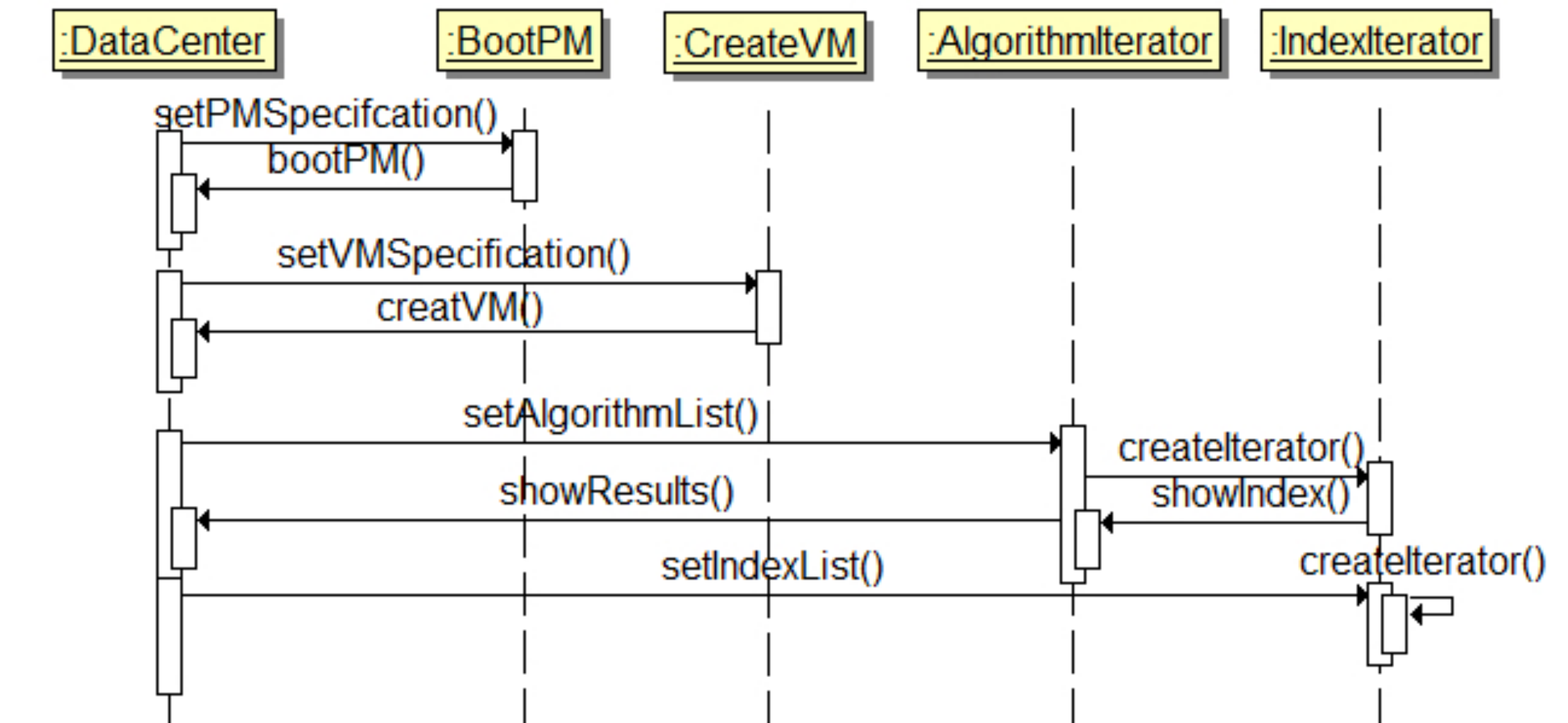}}
\caption{Sequence diagram of basic process}
\end{center}
\end{figure}

Fig.11 depicts the flow of communication among important FlexCloud entities. At the beginning of the simulation, a DataCenter entity sends necessary messages that BootPM entity needs to start PMs providing services in datacenter. CreateVM entity would also accept messages it needs to create VM requests. AlgorithmIterator and IndexIterator entities act to run scheduling algorithms and send calculated indices values back to DataCenter entity.

The communication flow described above is a basic flow in a simulated experiment. Some variations in this flow are possible depending on the scheduling process. For example, before bootPM() and createVM(), the message sent by a datacenter should be verified.

\section{Validation of FlexCloud}
To validate the accuracy of FlexCloud, we have designed some test cases to compare the theoretical results and simulation results. In this section, we use LS (List Scheduling), LPT(Longest Processing Time First), EDF(End-time Decreasing First) algorithms to compare theoretical and simulation results.
\begin{table}
\caption{Theoretical and simulation results comparison of LS algorithm}
\begin{center}
\begin{tabular}{|l|l|l|}
\hline LS Indices& Theoretical & Simulation
\\\hline
\hline Average Utilization & 0.5 & 0.5 \\
\hline Imbalance Degree & 0.0 & 0.0 \\
\hline Makespan & 0.5 & 0.5 \\
\hline Skew(makespan) & 1 & 1 \\
\hline Capacity\_makespan & 50 & 50 \\
\hline Skew(capacity\_makespan) & 1 & 1 \\
\hline
\end{tabular} \\
\end{center}
\end{table}

The test cases we designed are easily theoretically calculated and can reflect some general situations. For LS algorithm that always allocate a VM to the PM with the lowest load, we set that there are 100 PMs and 100 VMs requests both in the same types, the start-time of requests are ordered in increasing sequence, $1,2,3,\ldots, 100$, and all requests duration are 100 and require capacity is 0.5 of a PM. Since PMs number and VMs number are same in this case, LS algorithm works as Round-Robin algorithm, that means each PM would undertake a VM task. Then we calculate the values in theoretical way and simulation, same results have been observed and shown in TABLE 4.
\begin{table}
\caption{theoreticcal and simulation results comparison of LPT algorithm}
\begin{center}
\begin{tabular}{|l|l|l|}
\hline LPT Indices& Theoretical & Simulation
\\\hline
\hline Average Utilization & 0.505 & 0.505 \\
\hline Imbalance Degree & 0.0 & 0.0 \\
\hline Makespan & 1 & 1 \\
\hline Skew(makespan) & 1 & 1 \\
\hline Capacity\_makespan & 50.5 & 50.5 \\
\hline Skew(capacity\_makespan) & 1 & 1 \\
\hline
\end{tabular} \\
\end{center}
\end{table}

We also design a test case for LPT algorithm, an offline algorithm that VM requests can be reordered by processing time before they are allocated. In this case, there are 50 PMs and 100 VMs both in the same types, each request requires 0.5 capacity of a PM and starts at $1,2,3,\ldots,100$, and the durations of VMs are ordered in decrease order from 100 to 1 as $100,99,98,\ldots,1$. Same results have been observed and collected in TABLE 5.
\begin{table}
\caption{theoretical and simulation results comparison of EDF algorithm}
\begin{center}
\begin{tabular}{|l|l|l|}
\hline EDF Indices& Theoretical & Simulation
\\\hline
\hline Power Consumption & 250000 & 250000 \\
\hline Rejected Number & 10 & 10 \\
\hline Turned on PMs & 20 & 20 \\
\hline
\end{tabular} \\
\end{center}
\end{table}

For energy saving algorithm EDF, it should be noticed that comparison indices are different and requests are ordered by end-time. In this case, we set that there are 20 PMs and 50 VMs both in the same types, the start-times of VM requests are ordered in increasing as $1,2,3,\ldots,50$ and end-times are decreasing as $100,99,98,\ldots, 51$. Each VM requires 0.5 capacity of a PM. We adopt the energy saving model referred to section 3.4 and assume $P_{min} = 300$, $P_{max} = 500$. Same theoretical and simulation values have been collected in TABLE 6. Referring to the collected data in TABLE 5 and 6, the results show the correctness of FlexCloud.

\section{Evaluations}
In this section, we provide more performance evaluations for FlexCloud, including evaluations for different algorithms with basic and advanced settings.
\subsection{Basic Algorithm Performance Evaluations}

\begin{figure*}[!ht]
\begin{center}
{\includegraphics [width=1.0\textwidth, height=1.5in, angle=-0] {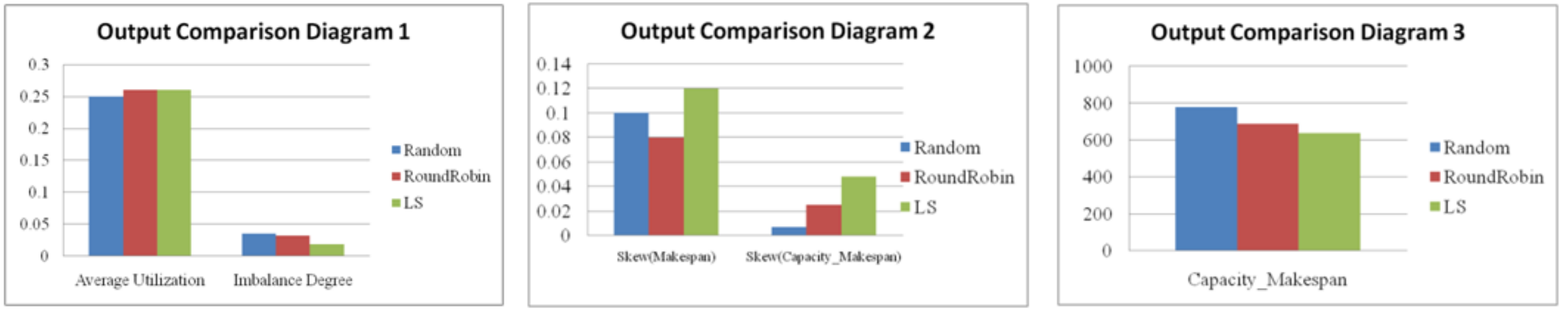}}
\caption{Output comparison Diagram}
\end{center}
\end{figure*}
To begin with, we compare scheduling algorithms performance with basic settings, and show the comparison diagrams generated by FlexCloud.
The related settings are as followings:\\
1) Algorithm type: is online load balancing;\\
2) PM specifications: using suggested specifications in Amazon EC2 shown in Table 2. PM type1 number is 50, type2 and type3 number are set as 0 to simplify simulation;\\
3) VM requests: using suggested specifications in Amazon EC2 as shown in Table 3 and 3, and requests are generated under Normal Distribution.
\\
4) Algorithms for comparison: Random, RoundRobin (R-R) and List Scheduling algorithm (LS, referring to section 3.3);\\
5) Indices for comparison: average utilization, imbalance degree, capacity\_makespan, skew of makespan, skew of capacity\_makespan.\\

FlexCloud provides several output formats for further analysis, like diagram outputs, text outputs or outputs in Excel file. The diagram output results are showed as bar chart presented in Fig. 12, which is composite of three small diagrams. As seen from these diagrams, it can be concluded that LS overwhelms the other two algorithms on imbalance degree, makespan, skew of makespan, and the skew of capacity\_makespan with the settings. It¡¯s easy to understand as LS algorithm dynamically allocates VM requests based on the PM loads while Random and RoundRobin algorithms do not collect real-time load information from PMs.
\subsection{Advanced Algorithm Performance Evaluations}
To extend performance evaluations, we also compare scheduling algorithms performance with advanced settings and collect the comparison data. The related settings are as following:\\
1) Algorithm type is offline load balancing;\\
2) PM specifications: using suggested specifications in Amazon EC2 shown in Table I. PMs with different numbers are considered. PMs numbers are varying from 15, 30, 60 to 240 and each type of PMs occupies about 1/3 of total PMs numbers;\\
3) VM requests: using suggested specifications in Amazon EC2 shown in Table II. We adopt the log data at Lawrence Livermore National Lab (LLNL) to reflect realistic data generation. The log contains months of records collected by a large Linux cluster and has characteristics consistent with our problem model. Each line of data in that log file includes 18 elements, while we only need the request-ID, start-time, duration and number of processors (capacity demands) in our simulation. We convert the units from seconds in LLNL log file into minutes, as we design 5 minutes to be a time slot length;\\
4) Algorithms for comparison: RoundRobin (R-R), Longest Processing Time first (LPT, referring to section 5), Post Migration Algorithm (MIG, referring to section 3.5), Capacity\_makespan Prepartition Algorithm (CMP, referring to section 3.5);\\
5) Indices for comparison: average utilization, imbalance degree, longest process time and capacity\_makespan.\\

\begin{figure*}[!ht]
\begin{center}
{\includegraphics [width=1.0\textwidth,height=1.5in, angle=-0] {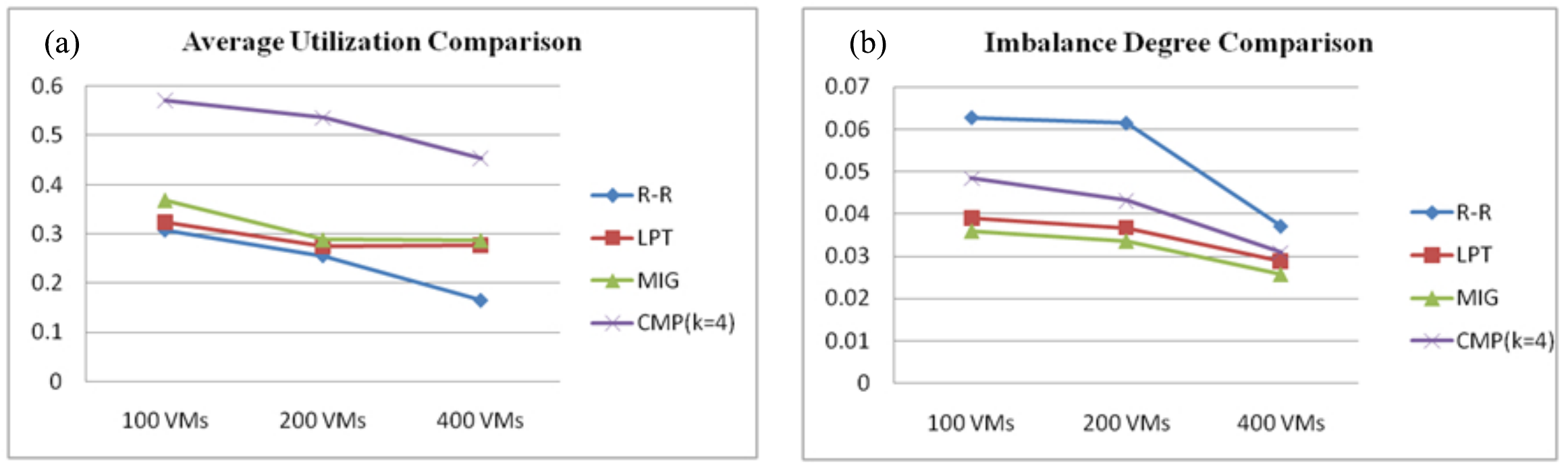}}
\caption{The offline algorithm comparison of average utilization (a) and imbalance degree (b) with LLNL trace}
\end{center}
\end{figure*}

\begin{figure*}[!ht]
\begin{center}
{\includegraphics [width=1.0 \textwidth, height=1.5in,angle=-0] {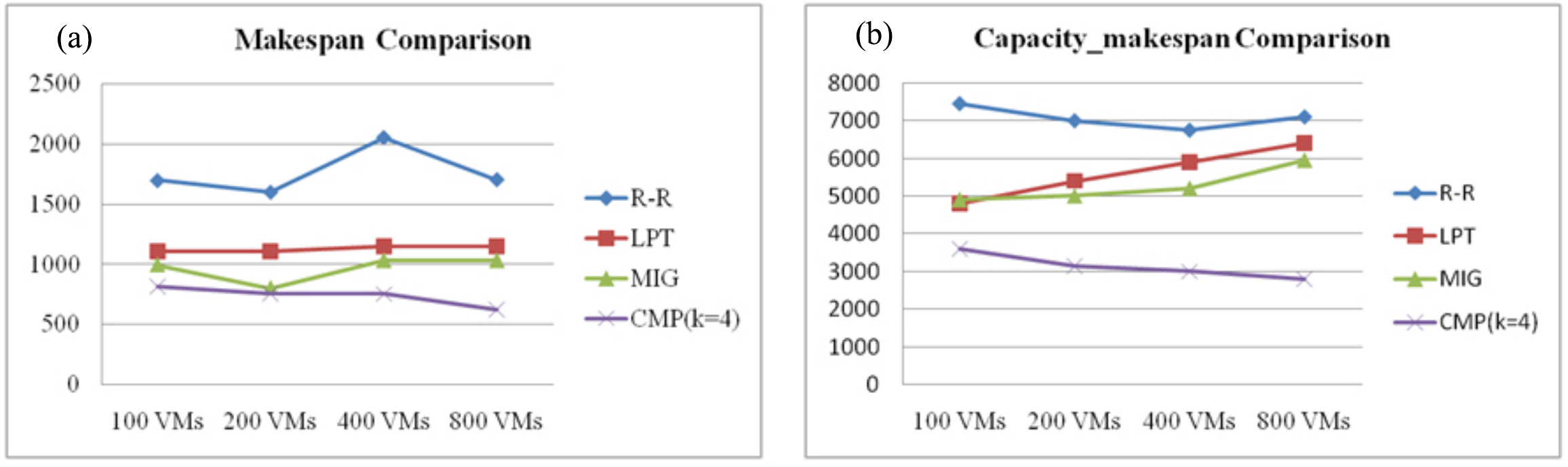}}
\caption{The offline algorithm comparison of makespan (a) and capacity\_makespan (b) with LLNL trace}
\end{center}
\end{figure*}

Fig.13 to Fig.14 show the average utilization, imbalance degree, makespan and capacity$\_$makespan comparison for different algorithms with LLNL data trace. From these figures, we can notice that CMP algorithm has better performance than other algorithms in average utilization, imbalance degree, makespan, capacity$\_$makespan. CMP algorithm has 10$\%$-20$\%$ higher average utilization than MIG and LPT, and 40$\%$-50$\%$ higher average utilization than Random-Robin (R-R). Prepartition algorithm has 10$\%$-20$\%$ lower average makespan and capacity$\_$makespan than MIG and LPT, and 40$\%$-50$\%$ lower average makespan and capacity$\_$makespan than R-R.

Besides the above evaluations, we also vary the partition number $k$ from 4, 8 to 10 to compare the load balance affects. Fig.15 presents imbalance degree 
of Capacity\_makespan Prepartition algorithm with different $k$ values. It's easy to understand that a larger $k$ value would produce a better load balance, which would lead to more partitions, and more partitions could achieve better load balance effects. It can be observed that whatever numbers of migrations to taken, Post Migration algorithm (MIG) just cannot achieve the same level of average utilization, makespan and capacity$\_$makespan as Capacity\_makespan Prepartition does.

\begin{figure}[!ht]
\begin{center}
{\includegraphics [width=0.5\textwidth, height=1.5in,angle=-0] {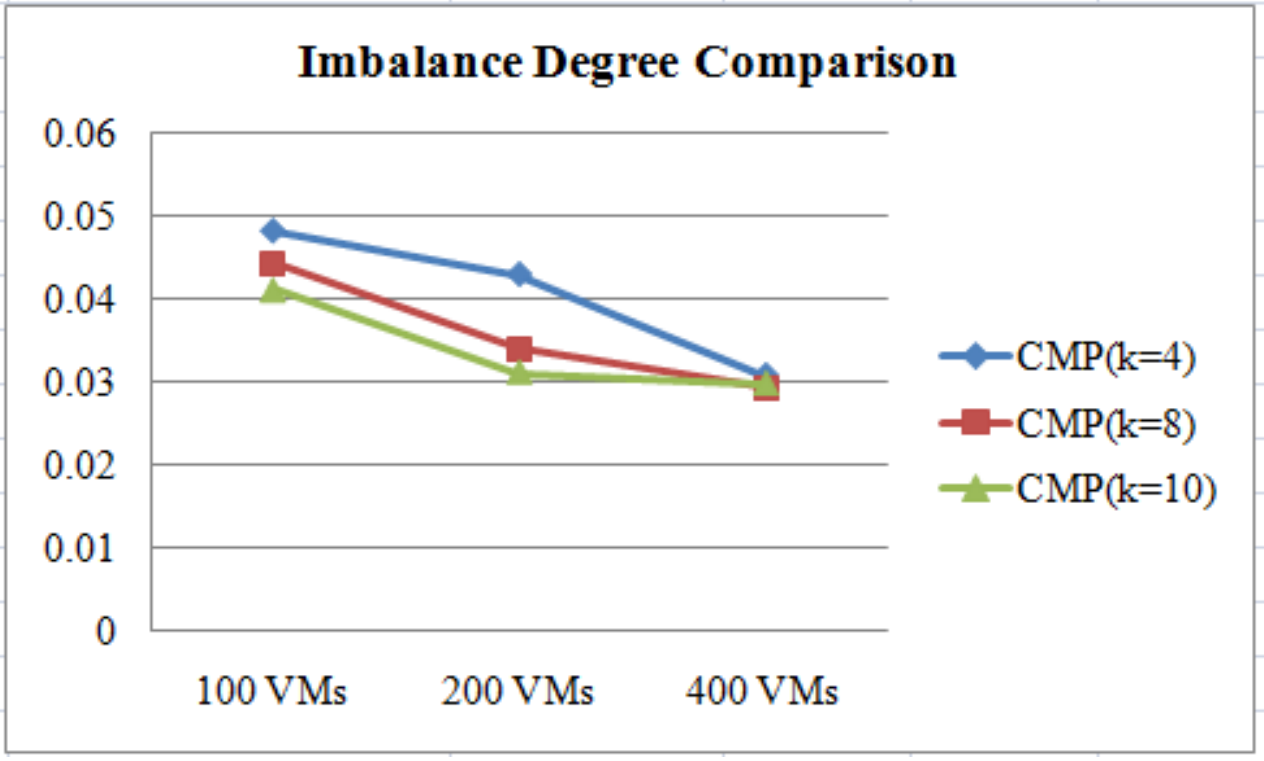}}
\caption{The comparison of Time Cost by varying $k$ values}
\end{center}
\end{figure}

\section{Conclusions and Further work}
In this paper, we introduce the FlexCloud, a novel simulator for performance evaluation of virtual machine allocation in Cloud data centers. It is flexible, scalable to simulate resource scheduling in cloud data centers. A complete simulation framework has been built and introduced.

There are a few research directions for extending the simulator:

\begin{itemize}
\item
Considering more scheduling algorithms. In FlexCloud, we already implemented load-balancing and energy-efficiency, other scheduling algorithms such as cost-oriented or reliability-oriented algorithms can be added in easily.
\item
Evaluate performance by datasets from real traces. Currently we are collecting data from real cloud applications, more evaluating results can be provided by real traces and benchmarks.
\item
Providing more visual outputs such as dashboards and logical view of different data centers and their resource usages. This information is very important for managers and operators to have.
\item
Considering more infrastructures, such as networking devices. Currently FlexCloud considers bandwidth requests and allocations. The network devices such as three-tire switches and routers distributed in different data centers are under modeling consideration.

\section*{Acknowledgment}
This research is partly sponsored by the National Natural Science Foundation of China (NSFC) (Grant Number:61150110486), and by Scientific Funding of Central University (2012-2014).
\end{itemize}

\vspace{-4em}
\begin{IEEEbiographynophoto}{Minxian Xu}
is a graduate student at the School of Software,
University of Electronic Science and Technology of China, Chengdu, under the
supervision of Dr. Tian.
\end{IEEEbiographynophoto}
\vspace{-4em}
\begin{IEEEbiographynophoto}{Wenhong Tian}
finished  his PhD degree in Computer Science, from North Carolina State University in 2007. Currently, he is an associate professor at University of Electronic Science and Technology of China. His research interests include dynamic design and resource management in random changing networks, Cloud computing, RFID middleware design, biocomputing. As the first author, he has published about 30 journal and conference papers in related research areas during recent years.
\end{IEEEbiographynophoto}

\vspace{-4em}
\begin{IEEEbiographynophoto}{Xinyang Wang}
is a graduate student at the School of Computer Science,
University of Electronic Science and Technology of China, Chengdu, under the
supervision of Dr. Tian.
\end{IEEEbiographynophoto}
\vspace{-4em}
\begin{IEEEbiographynophoto}{Qin Xiong}
is a graduate student at the School of Computer Science,
University of Electronic Science and Technology of China, Chengdu, under the
supervision of Dr. Tian.
\end{IEEEbiographynophoto}


\nocite{*}
\begin{thebibliography}{99}
\bibliographystyle{compj}
\bibliography{CRESS_ComputerJournal3}
\bibitem{IEEEhowto:Beloglazov}
A. Beloglazov, J.Abawajy, R. Buyya, Energy-Aware Resource Allocation Heuristics for Efficient Management of Data Centers for Cloud Computing Future Generation Computer Systems, Volume 28, Issue 5, May, pp.755-768, 2012.
\bibitem{IEEEhowto:Legrand}
A. Legrand, L. Marchal, and H. Casanova. Scheduling distributed applications: the SimGrid simulation framework. Proceedings of the 3rd IEEE/ACM International Symposium on Cluster Computing and the Grid, 2003.
\bibitem{IEEEhowto:Alberto}
A. Nunez, J. Vazquez-Poletti, A. Caminero et al., iCanCloud: A Flexible and Scalable Cloud Infrastructure Simulator, J. Grid Computing 10:185, C209, 2012
\bibitem{IEEEhowto:Amazon}
Amazon EC2, http://aws.amazon.com/ec2/
\bibitem{IEEEhowto:Singh}
A. Singh, M. Korupolu, D. Mohapatra, Server-Storage Virtualization: Integration and Load Balancing in Data Centers, in the proceedings of the 2008 ACM/IEEE conference on Supercomputing, pp.1-12, 2008.
\bibitem{IEEEhowto:Wickremasinghe}
B. Wickremasinghe et al., CloudAnalyst: A CloudSim-based Tool for Modelling and Analysis of Large Scale Cloud Computing Environments, Proceedings of the 24th IEEE International Conference on Advanced Information Networking and Applications (AINA 2010), Perth, Australia, April 20-23, 2010.
\bibitem{IEEEhowto:Dumitrescu}
C. L. Dumitrescu and I. Foster. GangSim: a simulator for grid scheduling studies. Proceedings of the IEEE International Symposium on Cluster Computing and the Grid (CCGrid 2005), Cardiff, UK, 2005.
\bibitem{IEEEhowto:Mastroianni}
C. Mastroianni, M. Meo, G. Papuzzo, Probabilistic Consolidation of Virtual Machines in Self-Organizing Cloud Data Centers. IEEE Transactions on Cloud Computing, vol.1, no.2, pp.215-228, Dec, 2013.
\bibitem{IEEEhowto:Economou}
D. Economou, S. Rivoire, C. Kozyrakis, P. Ranganathan, Full-System Power Analysis and Modeling for Server Environments, 2006. Stanford University / HP Labs Workshop on Modeling,Benchmarking, and Simu- lation (MoBS) June 18, 2006.

\bibitem{IEEEhowto:DMTF}
DMTF Cloud Management, http://www.dmtf.org/standards /cloud
\bibitem{IEEEhowto:Kliazovich}
D. Kliazovich, P. Bouvry, S.U. Khan, GreenCloud: a packet-level simulator of energy-aware, J SuperComput, 2010
\bibitem{Howell}
F. Howell and R. Mcnab. SimJava: A discrete event simulation library for java. Proceedings of the first International Conference on Web-Based Modeling and Simulation, 1998.
\bibitem{FlexCloud}
FlexCloud Project, http://sourceforge.net/projects/flexcloud/, 2014
\bibitem{IEEEhowto:Google}
Google App Engine, http://code.google.com/intl/zh-CN/appengine/
\bibitem{IEEEhowto:Zheng}
H. Zheng, L. Zhou, J. Wu, Design and Implementation of Load Balancing in Web Server Cluster System, Journal of Nanjing University of Aeronautics \& Astronautics, Vol. 38 No. 3 Jun. 2006.
\bibitem{IEEEhowto:ESL}
Hebrew University, Experimental Systems Lab, www.cs.huji.ac.il/labs/parallel/workload, 2013.
\bibitem{IEEEhowto:IBM}
IBM (2007) blue cloud, http://www.ibm.com/grid/
\bibitem{IEEEhowto:Doyle}
J. Doyle, R. Shorten, D. O'Mahony, Stratus: Load Balancing the Cloud for Carbon Emissions Control. IEEE Transactions on Cloud Computing, vol.1, no.1, pp.116-128, Aug, 2013.
\bibitem{IEEEhowto:Youseff}
L. Youseff, et al., 2008. Toward A Unified Ontology Of Cloud Computing, In the proceedings of Grid Computing Environments Workshop, GCE08, 2008.
\bibitem{IEEEhowto:Armbrust}
M. Armbrust, A. Fox, R. Griffith, A. Joseph, R. Katz, A. Konwinski, G. Lee, D. Patterson, A. Rabkin, I. Stoica, M. Zaharia. Above the Clouds: A Berkeley View of Cloud computing. Technical Report No. UCB/EECS- 2009-28, University of California at Berkley, USA, Feb. 10, 2009.
\bibitem{IEEEhowto:Xu2012}
M. Xu, W. Tian, An Online Load Balancing Schduling Algorithm for Cloud Data Centers Considering Real-time Multi-dimensional Resource, IEEE International Conference on Cloud Computing and Intelligence Systems, 2012.
\bibitem{IEEEhowto:Microsoft}
Microsoft (2008) Windows Azure, http://www.microsoft.com/windowsazure.
\bibitem{IEEEhowto:Buyya}
R. Buyya and M. Murshed. GridSim: A Toolkit for the Modeling and Simulation of Distributed Resource Management and Scheduling for Grid Computing. Concurrency and Computation: Practice and Experience, 14(13-15), Wiley Press, Nov-Dec., 2002.
\bibitem{IEEEhowto:Buyy2}
R.Buyya, R. Ranjan and R. N. Calheiros, Modeling and Simulation of Scalable Cloud Computing Environments and the CloudSim Toolkit: Challenges and Opportunities, Proceedings of the 7th High Performance Computing and Simulation Conference (HPCS 2009, ISBN: 978-1-4244-4907-1, IEEE Press, New York, USA), Leipzig, Germany, June 21 - 24, 2009.
\bibitem{IEEEhowto:Buyya3}
R.Buyya, C. S. Yeo, S. Venugopal, J. Broberg, and I. Brandic. Cloud Computing and Emerging IT Platforms: Vision, Hype, and Reality for Delivering Computing as the 5th utility. Future Generation Computer Systems, 25(6): 599-616, Elsevier Science, Amsterdam, The Netherlands, June 2009.
\bibitem{IEEEhowto:Prodan}
R. Prodan and M. Wieczorek, Bi-criteria scheduling of scientific grid workflows, IEEE Trans. Autom. Sci. Eng., vol. 7, no. 2, pp. 364376, Apr 2010.
\bibitem{IEEEhowto:Guerout}
T. Guerout, T. Monteil, G. D. Costa et al., Energy-aware simulation with DVFS, Simulation Modelling Practice and Theory, 39. pp 76-91, Dec 2013.
\bibitem{IEEEhowto:Wood}
T. Wood, et. al., Black-box and Gray-box Strategies for Virtual Machine Migration in the proceedings of Symp. on Networked Systems Design and Implementation (NSDI), 2007.
\bibitem{IEEEhowto:Zhang}
W. Zhang, Research and Implementation of Elastic Network Service, PhD dissertation, National University of Defense Technology, China (in Chinese) 2000.
\bibitem{IEEEhowto:Tian1}
W. Tian. Adaptive Dimensioning of Cloud Datacenters, in the proccedings of IEEE the 8th International Conference on Dependable, Autonomic and Secure Computing (DASC-09), Chengdu, China, December 12-14, 2009.
\bibitem{IEEEhowto:Tian2}
W. Tian, Y. Zhao, Y. Zhong, M. Xu, C. Jing, A dynamic and integrated load-balancing scheduling algorithm for Cloud data centers, In the proceedings of CCIS 2011, Beijing, China. Sept. 2011.
\bibitem{Tian2014}
W. Tian, M. Xu, Y. Chen, Y. Zhao, Prepartition:~A New Paradigm for the Load Balance of Virtual Machine Reservations in Data Centers, To appear in the proceedings of the IEEE International Conference on Communications (ICC), 2014.

\end{thebibliography}
\end{document}